\documentclass{emulateapj}

\usepackage{graphicx}
\shorttitle{The DENIS\,1048$-$3956 magnetosphere}
\shortauthors{Ravi et al.}

\begin{document}

\title{The magnetosphere of the ultracool dwarf DENIS\,1048$-$3956}

\author{V. Ravi,\altaffilmark{1,2}}
\affil{Research School of Astronomy and Astrophysics, ANU, Mt. Stromlo Observatory, 
Weston ACT 2611, Australia}

\author{G. Hallinan,\altaffilmark{3,4}}
\affil{National Radio Astronomy Observatory, 520 Edgemont Road, Charlottesville, VA 22903, USA}

\author{G. Hobbs,}
\affil{CSIRO Astronomy and Space Sciences, Australia Telescope National 
Facility, PO Box 76, Epping NSW 1710, Australia}

\and

\author{D. J. Champion}
\affil{Max-Planck-Institut f\"{u}r Radioastronomie, Auf dem H\"{u}gel 69, 53121, Bonn, 
Germany}

\altaffiltext{1}{CSIRO Astronomy and Space Sciences, Australia Telescope National 
Facility, PO Box 76, Epping NSW 1710, Australia}
\altaffiltext{2}{Present address: School of Physics, University of Melbourne, Parkville, VIC 3010, Australia}
\altaffiltext{3}{Department of Astronomy, University of California, Berkeley, CA 94720, USA}
\altaffiltext{4}{G. Hallinan is a Jansky Fellow of the National Radio Astronomy Observatory.}

\begin{abstract}
Ultracool dwarfs, the least-massive contributors to the stellar mass function, exhibit striking magnetic 
properties that are inconsistent with trends for more massive stars. Here, we present the widest-band 
radio observations to date of an ultracool dwarf, DENIS-P\,J104814.9$-$395604, in 
four 2\,GHz bandwidths between wavelengths of 1\,cm and 10\,cm. These data were obtained with the 
Australia Telescope Compact Array using the new Compact Array Broadband Backend instrument. We detected 
a stable negatively-sloped power-law spectrum in total intensity, with spectral index $\alpha=1.71\pm0.09$. 
Circular polarization fractions between 0.25 and 0.4 were found at the low-frequency end of our detection band. 
We interpret these results as indicative of gyrosynchrotron emission. We suggest that the radio emission originates 
from beyond the co-rotation radius, $R_{C}$, of the star. Adopting this model, we find $R_{C}$ between $1.2-2.9R_{*}$, 
and a non-thermal electron density and magnetic field strength between $10^{5}-10^{7.2}$\,cm$^{-3}$ and $70-260$\,G 
respectively at $R_{C}$. The model accounts for the violation of the G\"{u}del-Benz relation between X-ray and radio 
luminosities of low-mass stars by DENIS-P\,J104814.9$-$395604.

\end{abstract}


\section{Introduction}

Ultracool dwarfs (UDs), with spectral types M7 and cooler, are the coolest stars and include the 
least massive hydrogen-burning objects and brown dwarfs \citep{bs93}. The interiors of ultracool dwarfs 
should be fully convective. Observations of magnetic activity tracers \citep[e.g.,][]{bbf+10}, direct 
polarimetric magnetic field strength measurements \citep[e.g.,][]{rb07} and magnetic tomography 
\citep{mdp+08,dmp+08,mdp+10} have revealed vastly different magnetic properties 
for such fully convective stars compared to more massive solar-type stars.

X-ray and radio emission are tracers of magnetic activity in cool stars. 
X-ray emission is associated with thermal bremsstrahlung from coronas \citep{rgv85}, 
possibly heated through thermal dissipation of accelerated, energetic electrons \citep{gb93}. 
These electrons also radiate at radio wavelengths, as evinced by a fixed ratio of X-ray to radio luminosities 
for several classes of active cool stars \citep[the G\"{u}del-Benz relation,][]{gb93}. For radio frequencies 
between 5\,GHz and 8\,GHz, the radio luminosities per unit frequency, $L_{R}$, and the X-ray luminosities, 
$L_{X}$, are related as $L_{X}/L_{R}\sim10^{15.5\pm0.5}$\,Hz. This relation appears largely independent of a 
variety of stellar properties, including rotation for periods greater than $\sim12$ hours, binarity, 
spectral class and photospheric activity, and applies to both variable and quiescent emission 
phenomena. Phenomenologically, the G\"{u}del-Benz relation implies that magnetic activity, as manifested 
in coronal heating, results in a non-thermal electron density that is proportional to the degree of activity.

Established trends for cool stars linking magnetic activity, magnetic field strengths and 
structures and stellar rotation do not apply to UDs. The ratios of $L_{X}$ to the bolometric 
luminosities, $L_{bol}$, for UDs are below those expected from trends for more massive cool stars \citep{bbf+10}. Remarkably, 
the radio luminosities exhibit an opposite trend, increasing relative to $L_{bol}$ for UDs \citep{b06,bbf+10}. 
The sample of radio-loud UDs includes late-M dwarfs and L dwarfs; no T dwarf has 
yet been detected at radio wavelengths \citep[e.g.,][]{b06}. All radio-loud UDs violate the 
G\"{u}del-Benz relation. For their sample of radio-loud UDs, 
\citet{bbf+10} find $log(L_{X}/L_{R})\sim14$ for late-M dwarfs, and $log(L_{X}/L_{R})\sim12$ for cooler dwarfs. 
No clear trends have been identified between the radio luminosities of UDs and stellar properties, such as rotation 
and magnetic field strength \citep{bbf+10}. 

Radio emission from UDs is variable on timescales of years, hours and minutes. Some UDs have radio lightcurves 
that are periodic on the rotation periods of a few hours \citep{had+06,hbl+07,had+08,brp+09}. These 
lightcurves are either characterized by modulations \citep[e.g.,][]{had+06} or by short duty-cycle 
peaks lasting for a few minutes \citep[e.g.,][]{hbl+07}. Other UDs exhibit isolated flares when otherwise radio-loud, 
also on few-minute timescales \citep[e.g.,][hereafter BP05]{bp05}. The quickly-varying emission is generally 100$\%$ 
circularly-polarized \citep[e.g.,][]{had+08}. In addition, the types of radio emission observed from 
UDs changes on timescales of years \citep{oph+09}, varying between undetectable, quiescent and 
periodic. The variability of UD radio emission 
characteristics makes it hard to identify unbiased radio-loud samples for population studies.

Radio observations of UDs provide significant insight into conditions in UD magnetospheres. In this Letter, 
we present the widest-band radio observations yet reported for the UD DENIS-P\,J104814.9$-$395604 (hereafter DENIS1048), 
using the new Compact Array Broadband Backend \citep{fw02} at the 
Australia Telescope Compact Array \citep[ATCA,][]{fbw92}. We propose a magnetospheric model which accounts 
for the violation of the G\"{u}del-Benz relation. This model could provide interesting insights into the 
magnetic field and plasma environments of these enigmatic stars.

\section{Observations and data analysis}

\begin{deluxetable*}{ccccc}
\tabletypesize{\scriptsize}
\tablecaption{Details of observations.}
\tablewidth{0pt}
\tablehead{
\colhead{Observing date} & \colhead{Center frequencies (GHz)} & \colhead{Start \& end times (UT)} & 
\colhead{Synthesized beam sizes}
}
\startdata
2009 August 9, 10 & 18, 24 & 21:10, 06:10 & 1.2\arcsec$\times$0.76\arcsec, 0.90\arcsec$\times$0.57\arcsec \\
2009 August 11 & 18, 24 & 03:25, 07:45 & 3.1\arcsec$\times$0.80\arcsec, 2.34\arcsec$\times$0.60\arcsec \\
2009 August 15 & 5.5, 9 & 02:30, 07:20 & 7.5\arcsec$\times$2.0\arcsec, 4.6\arcsec$\times$1.2\arcsec
\enddata
\end{deluxetable*}

\begin{figure*}[h!]
\centering
\includegraphics[angle=-90,scale=0.55]{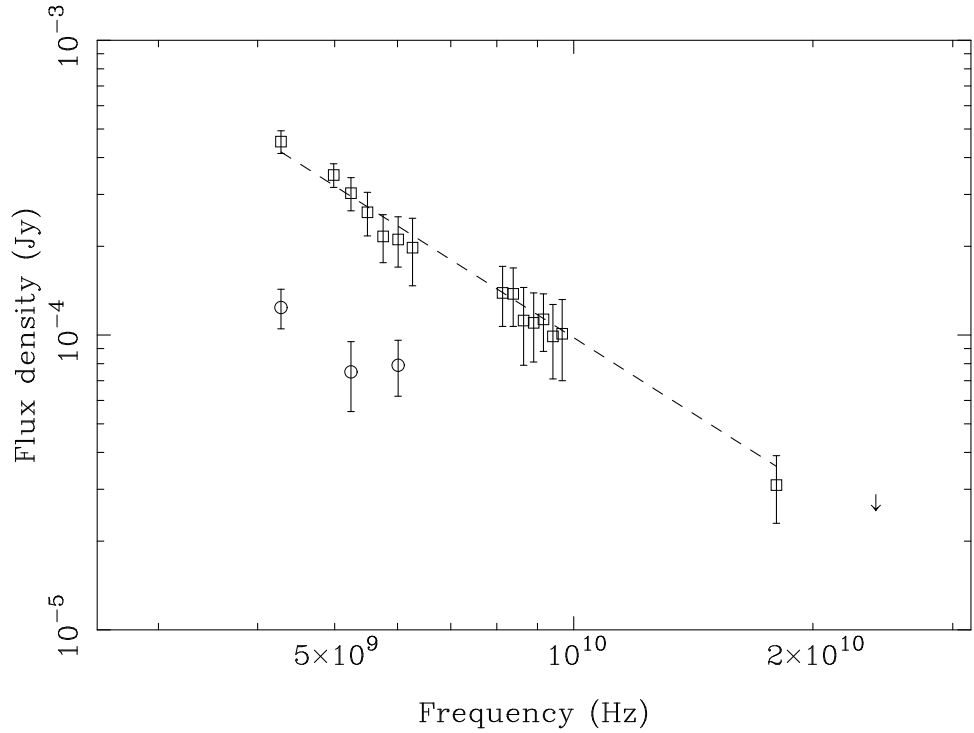}
\caption{The radio spectrum of DENIS1048 obtained with the ATCA between 2009 August 11 and 15. The squares are 
the Stokes I measurements, the circles are the Stokes V measurements and the arrow indicates the $3\sigma$ 
upper limit we place for the 24\,GHz emission. The dashed line is the best-fit power law 
($S(\nu)\propto\nu^{-\alpha}$) to the Stokes I spectrum, with index $\alpha=1.71\pm0.09$.}
\end{figure*}

The target source, DENIS1048, was one of seven Southern late-M and L dwarfs 
observed by BP05 with the ATCA in the 3\,cm and 6\,cm bands. BP05 reported a quiescent flux density of 
0.14$\pm$0.04\,mJy at 6\,cm, as well as a 4$-$5 minute flare in each band, separated by 
$\sim$10 minutes with a peak flux density of 30\,mJy at 3\,cm. DENIS1048 
\citep[spectral classification M8.5,][]{hsb+04} was identified as a UD in the DENIS survey \citep{edc+97} by 
\citet{dfm+01}, and, at a distance of 4.00$\pm$0.03\,pc \citep{cmj+05}, is one of the closest known stars. 
A recent spectroscopic study by \citet{mpb+10} shows that it is unlikely to be a brown dwarf. 
\citet{fs04} reported a large optical flare and a fast projected rotation velocity of $v\sin i=25\pm2$\,km\,s$^{-1}$. 
While H$\alpha$ emission was detected by \citet{dfm+01}, \citet{sl04} found no X-ray emission with an upper 
limit of $2\times10^{26}$\,erg\,s$^{-1}$ from the ROSAT All-Sky Survey catalogue. \citet{rb10} found an average 
line-of-sight magnetic field strength of $2300\pm400$\,G using measurements of Zeeman broadening in FeH absorption 
lines. 

We observed DENIS1048 with the ATCA on 2009 August 10 and 11 in the 1.2\,cm band, and simultaneously in the 
3 and 6\,cm bands on 2009 August 15. The six 22-metre ATCA antennas were placed in an extended configuration in 
order to maximise point-source sensitivity. Baseline lengths ranged between 300\,m and 6000\,m, corresponding to 
resolutions between approximately 40\arcsec~and 2\arcsec~at 6\,cm. 
Visibility measurements for all baselines were recorded in two 2.048\,GHz bands 
per Stokes polarization with 1\,MHz frequency resolution. The visibilities were integrated 
over 10\,s intervals. Details of the observations are given in Table 1. 

We reduced the data using the MIRIAD software package \citep{stw95}. Standard calibrations 
were performed using observations of the ATCA primary calibrator PKS\,B1934$-$638 on each day, and 
frequent observations of a radio galaxy $-$ PKS\,B1104$-$445 $-$ separated by 6$^{\circ}$ from 
DENIS1048. Multi-frequency synthesis total-intensity images were produced in sub-bands of 256\,MHz 
for the 6\,cm and 3\,cm observations, and in each 2\,GHz sub-band for the 1.2\,cm observations. 
We detected DENIS1048 as a point source in all images, except that formed from the 24\,GHz data. 
The measured position of right-ascension: 10h, 48m, 13.58s ($\pm0.03$s), declination: -39$^{\circ}$, 56', 
16.0\arcsec~($\pm0.5$\arcsec) is offset from the 2MASS position of DENIS1048 \citep{c+03} by 15.6\arcsec, which 
corresponds to the known proper motion \citep{dhc05}. The flux density of DENIS1048 
was measured in each sub-band by fitting the restoring Gaussian beams to the images. 
The beams were at position angles of $19^{\circ}$ for the 6\,cm and 3\,cm data, and $2^{\circ}$ for the 1.2\,cm data, 
The rms noise levels, $\sigma$, in images made from each sub-band ranged between $30-50$\,$\mu$Jy for the 6\,cm data, and between 
$25-35$\,$\mu$Jy for the 3\,cm data. For both the 18\,GHz and 24\,GHz, $\sigma=8$\,$\mu$Jy. 

We present the resulting spectrum of DENIS1048 in Figure 1. The Stokes I measurements of the flux density of 
DENIS1048, $S(\nu)$, at various frequencies $\nu$ are an excellent fit to a power-law, $S(\nu)\propto\nu^{-\alpha}$, 
where $\alpha=1.71\pm0.09$.

\begin{figure*}[h!]
\centering
\includegraphics[scale=0.55]{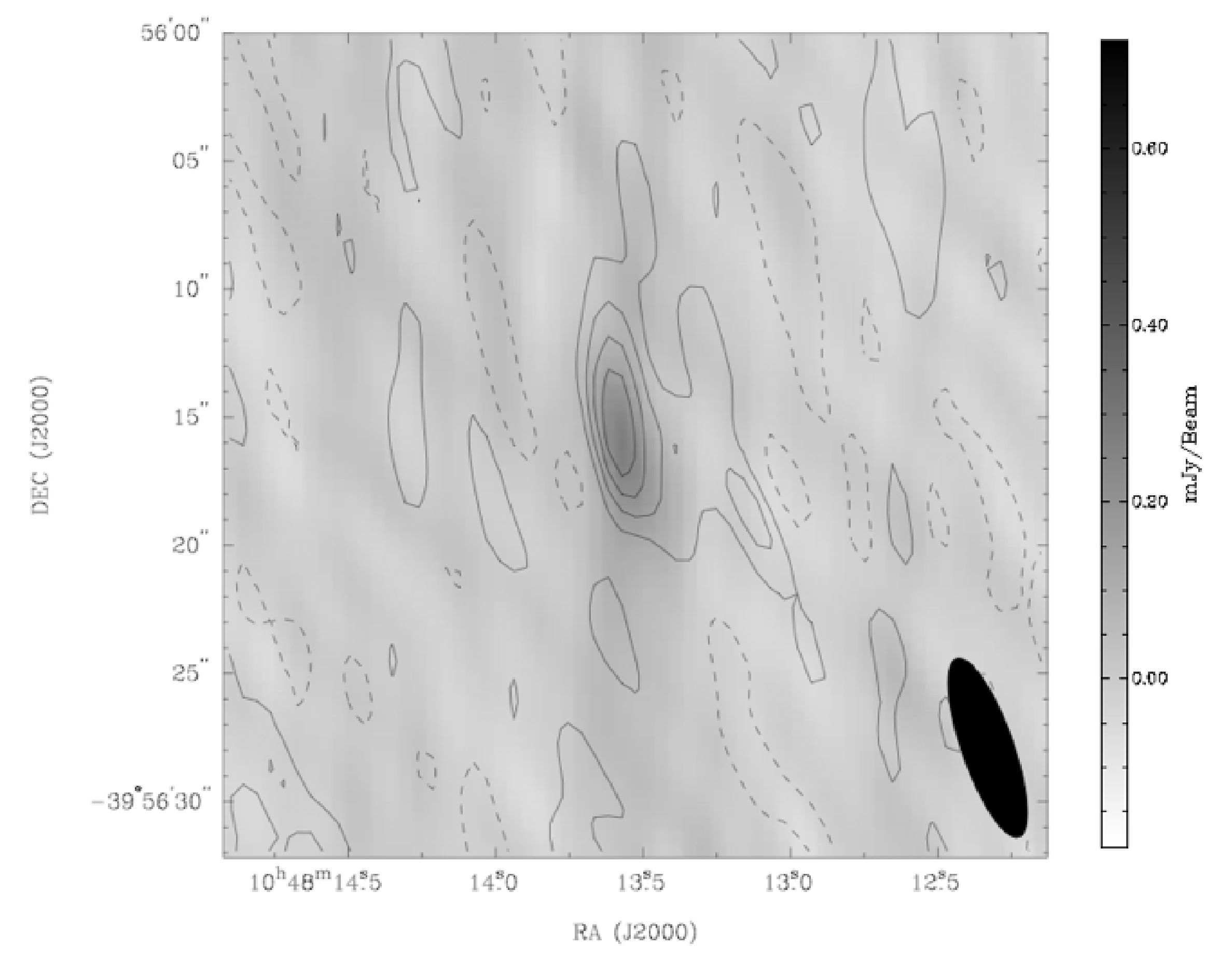}
\caption{Image of the Stokes I (greyscale) $\&$ Stokes V (contours) emission from DENIS1048 made using 
data at frequencies between 4.576 $\&$ 4.988\,GHz. These data were obtained on 2009 August 15 using the ATCA, 
with a maximum baseline length of 6\,km. The beam used to make the image, shown as a filled ellipse in the 
bottom-right corner, had a size of 8.7\arcsec$\times$2.3\arcsec~at a position angle of $19^{\circ}$. The Stokes V 
contour levels are at $-1\sigma$ (dashed), $1\sigma,\;2\sigma,\;3\sigma,\;\&\;4\sigma$, where $\sigma=25\,\mu$Jy. 
The maximum Stokes I intensity in the field is 370\,$\mu$Jy.}
\end{figure*}

A similar process was also applied to the Stokes Q, U and V data. While DENIS1048 was not found to have any detectable 
Stokes Q or U emission, we detected Stokes V emission at frequencies up to 6.5\,GHz by combining data in 
multiple sub-bands. Circular polarization fractions ranging between 0.25 and 0.4 were found in the 6\,cm band, and 
3$\sigma$ upper limits of 0.2 were placed on the linear polarization fractions. The Stokes V flux density measurements 
are also plotted in Figure 1. An image of the data recorded between 4.576\,GHz and 4.988\,GHz, with Stokes V contours 
overlayed on a Stokes I greyscale image, is shown in Figure 2. No significant short-timescale amplitude excursions 
or periodicities were detected in any sub-band in either the Stokes I or V data. 

\section{The radio emission mechanism}

We consider two possible radio emission mechanisms for DENIS1048: gyrosynchrotron emission and 
electron-cyclotron maser (ECM) emission. The spectral shape and amplitude are clearly inconsistent with thermal 
emission. Gyrosynchrotron and ECM mechanisms are both associated with mildly relativistic, 
non-thermal electron populations, with energies $>\sim20$\,keV. Whereas gyrosynchrotron emission 
is caused by incoherently radiating non-thermal electrons propagating along magnetic field lines, 
ECM emission is coherent, and requires these electrons to have an anisotropic pitch-angle distribution \citep{d85}. 
Gyrosynchrotron emission is characterized by a negatively-sloped power-law spectrum at high frequencies, 
corresponding to optically thin emission, and a positively-sloped spectrum at lower frequencies corresponding to 
optically thick emission. ECM emission, however, occurs at frequencies corresponding to low harmonics of 
the local cyclotron frequency, and does not have a characteristic spectral shape \citep{m09}. 

Time-variable emission from UDs is up to 100\% circularly polarized and tightly beamed, and is hence 
interpreted as ECM emission \citep[e.g.,][]{had+08}. Gyrosynchrotron emission has been hypothesized by a 
variety of authors for radio-loud UDs in quiescent states \citep[e.g., BP05;][]{ohb+06}. 

The observed radio properties of DENIS1048 are more suggestive of gyrosynchrotron emission than ECM 
emission for three reasons:
\begin{itemize}
\item The negatively-sloped power-law spectrum we observe is expected of optically-thin gyrosynchrotron emission, 
whereas no particular spectral shape is uniquely identified with ECM emission.
\item If the 1.2\,cm emission were caused by an ECM mechanism, the emission would need to originate 
in a stable region with a magnetic field strength of approximately 6.4\,kG to be consistent with 
emission at the fundamental cyclotron frequency. Unless the line of sight is nearly aligned with the 
rotation axis, such a region must also cover a significant fraction of the stellar surface, which is unlikely 
\citep{rb10}.
\item The lack of significant modulation is marginally inconsistent with the tight beaming of ECM emission.
\end{itemize}
While we cannot conclusively rule out an ECM emission mechanism, we interpret the radio emission we observe from DENIS1048 as 
optically thin gyrosynchrotron emission. In this interpretation, the large observed Stokes V fractions are 
consistent with a strong line-of-sight magnetic field component. Furthermore, the number of radiating 
electrons, $N(E)$, per unit energy, $E$, can be written as a power law, $N(E)\propto E^{-\delta}$, where 
$\delta=(1.22-\alpha)/0.9$ \citep{d85}. We find $\delta=3.26\pm0.09$, consistent with expected values between 
$-2$ and $-7$ for gyrosynchrotron-emitting electrons \citep{d85}. 

\section{A model for the magnetosphere}

Assuming $L_{X}<2\times10^{26}$\,erg\,s$^{-1}$ after \citet{sl04} for DENIS1048, the G\"{u}del-Benz relation 
implies an average radio flux density between 5\,GHz and 8\,GHz of less than 42\,$\mu$Jy. At the lowest observing 
frequency, the measured flux density is an order of magnitude greater.

We propose a model for the magnetosphere of DENIS1048 that accounts for the violation of the G\"{u}del-Benz relation. 
For samples of fast-rotating M dwarfs, an observed decrease in $L_{X}/L_{bol}$ with rotation period 
\citep{bbf+10,jjb+11} is interpreted as evidence for the decoupling of hot coronal plasma beyond a co-rotation 
radius, $R_{C}$. We hypothesize that this effect does not reduce $L_{R}/L_{bol}$. We further suggest that:
\begin{enumerate}
\item $R_{C}$ represents the Alfv\`{e}n radius of the star, and the magnetic field structure beyond $R_{C}$ is 
radial approximately and dominated by the outflow of non-thermal electrons.
\item Electron acceleration occurs at $R_{C}$, possibly through magnetic reconnection \citep{zy09}. 
\end{enumerate}
In this model, the gyrosynchrotron emission we observe originates from non-thermal electrons streaming radially 
outwards from $R_{C}$. A radial magnetic field structure in this region is justified 
by the lack of significant variability in the radio lightcurve.

The radio emission is characterized by three parameters: $R_{C}$, the total non-thermal electron 
density, $N_{e}$, at $R_{C}$, and the radial magnetic field strength, $B_{C}$, at $R_{C}$. We attempted to uniquely 
measure these parameters through a fit to the measured spectrum of DENIS1048, using expressions from \citet{d85} for the 
total and circularly-polarized intensities of gyrosynchrotron emission. A large number of parameter combinations 
were found to fit the data. Two assumptions, however, allowed the free parameters to be constrained within the 
ranges given in Table 2. First, we assumed that the spectral peak of the gyrosynchrotron emission \citep{d85} 
did not lie within the spectral band, as justified by the regularity of the observed power-law spectrum. We also limited 
$B_{C}<2700/(R_{C}-R_{*})^{3}$\,G in the dipole approximation, where 2700\,G is the 
upper limit on the surface magnetic field strength placed by \citet{rb10}. The stellar radius, $R_{*}$, is further 
assumed \citep[after, e.g.,][]{bag95,bhl+01,bp05,oph+09} to be equivalent to a Jupiter radius. 

\begin{deluxetable}{ccc}
\tabletypesize{\scriptsize}
\tablecaption{Constrained ranges for the model parameters.}
\tablewidth{0pt}
\tablehead{
\colhead{$R_{C}/R_{J}$} & \colhead{$N_{e}$ (cm$^{-3}$)} & \colhead{$B_{C}$ (G)} 
}
\startdata
1.2 to 2.9 & $10^{5}$ to $10^{7.2}$ & 70 to 260
\enddata
\end{deluxetable}

\section{Conclusions and future work}

We have utilised the unprecedented frequency coverage of the the upgraded ATCA to characterize the radio spectrum of 
the UD DENIS1048. Between 4.5\,GHz and 24\,GHz, the spectrum follows a strikingly regular power-law 
with index $\alpha=1.71\pm0.09$. This spectral shape, the lack of time-variability, and circular polarization 
fractions between 0.25$-$0.4 in the 6\,cm band, are suggestive of a gyrosynchrotron mechanism for the radio emission. 

Further work is required to improve and test the model we present. An observation of a peak frequency in the 
spectrum of DENIS1048 would break the degeneracy between the free parameters. It is possible that the model 
predicts correlations between radio emission characteristics of UDs and the stellar parameters. Such 
correlations are yet to be investigated. Finally, the model does not account for radio emission other than from a 
gyrosynchrotron mechanism, or for time-variable emission. Efforts are under way to model UDs in such states 
\citep[e.g.,][]{yhd+10}.

\acknowledgements

We thank the ATCA staff and duty astronomers for their valuable assistance with the observations reported here. 
The Australia Telescope Compact Array is part of the Australia Telescope which is funded by the Commonwealth of 
Australia for operation as a National Facility managed by CSIRO. This research has made use of the SIMBAD database, 
operated at CDS, Strasbourg, France. GH is supported by an Australian Research Council QEII Fellowship (project \#DP0878388). 
The National Radio Astronomy Observatory is a facility of the National Science Foundation operated under cooperative 
agreement by Associated Universities, Inc.

\end{document}